\documentclass[fleqn,10pt]{wlscirep}
\usepackage[utf8]{inputenc}
\usepackage[T1]{fontenc}
\usepackage{subfigure,color}
\usepackage[sort&compress,square,numbers]{natbib}
\title{Huygens metasurface supporting quasi-bound states in the continuum for terahertz gas sensing}

\author[+,*]{Jose Antonio \'Alvarez-Sanchis}
\author[+]{Borja Vidal}
\author[+]{Ana D\'iaz-Rubio}
\affil[+]{Nanophotonics Technology Center, Universitat Polit\`ecnica de Val\`encia, Valencia, Spain.}
\affil[*]{jaalvsa1@ntc.upv.es}

\begin{abstract}
We investigate a terahertz (THz) gas sensing platform based on all-dielectric metasurfaces that support quasi-bound states in the continuum (quasi-BIC) with both electric and magnetic dipole resonances. The structure is designed to achieve the first Kerker condition, minimizing backscattering and maximizing light-matter interaction, which significantly enhances the sensitivity of the sensor. By optimizing structural parameters, this metasurface selectively resonates at characteristic absorption frequencies of target gases, facilitating detection even at low concentrations. We validate the approach using two gases with strong but distinct THz absorption profiles: hydrogen cyanide (HCN) and sulfur dioxide (SO$_2$). Furthermore, the free-standing design maximizes gas interaction on both sides of the metasurface, eliminating substrate-induced losses and enabling a reduced physical footprint. Our findings indicate that this metasurface outperforms standard THz sensing approaches in terms of compactness and sensitivity per path length unit, obtaining the same detection threshold as sensing in free space with a path length between 2 and 3 orders of magnitude shorter, underscoring its potential for industrial applications where the available space for sensing can be limited.
\end{abstract}
\begin{document}

\flushbottom
\maketitle
%
%
\thispagestyle{empty}

\section*{Introduction}
Gas sensing is essential across a variety of critical applications, including quality control in industrial processes \citep{Allen_Diode_1998}, environmental pollutant monitoring \citep{McKercher_Characteristics_2017}, and hazard detection in safety-sensitive settings \citep{Maleki_Detection_2015}. Among the numerous gas sensing techniques, spectroscopic methods are especially valued for their precision, selectivity, and non-invasive nature \citep{Hodgkinson_Optical_2012}. Many gases exhibit distinctive absorption peaks within the mid-infrared (mid-IR) and terahertz (THz) spectral regions, making these wavelengths suitable for gas identification through methods such as THz time-domain spectroscopy \citep{Mittleman_Gas_1998} and Fourier-transform infrared spectroscopy \citep{Li_Monitoring_2002}. To enhance the sensitivity of these techniques and enable detection of gases at lower concentrations while maintaining a minimal physical footprint, several advanced approaches have been developed. These include the use of resonant cavities that increase the effective optical path \citep{Wang_Dense_2022, Elmahle_THz_2023} , microporous membranes that absorb gas molecules and concentrate them along the optical path \citep{You_Terahertz_2015}, as well as photonic crystals \citep{Qin_Terahertz_2019, Shi_Highly_2018, Chen_Terahertz_2014} and metasurfaces based on surface-enhanced infrared absorption (SEIRA) spectroscopy \citep{Chang_All_2018, Lee_Advancements_2024}.

Metasurfaces, in particular, have demonstrated significant potential for material characterization due to the field enhancement at resonant frequencies, which intensifies light-matter interactions. Different metasurface configurations can be harnessed for sensing, including metallic split-ring resonators \citep{Driscoll_Tuned_2007, Hu_Study_2015}, metasurfaces employing extraordinary optical transmission \citep{Jauregui_THz_2018}, and dielectric metasurfaces that leverage quasi-bound states in the continuum (quasi-BIC) resonances \citep{Wang_All_2021, Tittl_Imaging_2018}. In metasurface-based sensing, two primary approaches are typically employed: (i) monitoring the resonance frequency, which shifts based on the sample’s refractive index \citep{Beruete_Terahertz_2020}, and (ii) identifying materials by detecting specific absorption peaks associated with the sample’s absorption spectrum \citep{Klingbeil_Temperature_2007, Fischer_Chemical_2007}. Although resonance frequency monitoring can achieve high sensitivity, it often lacks specificity in complex gas mixtures where multiple gases share similar refractive indices \citep{Edmonson_Gas_1957}.
Absorption-based sensing, by contrast, utilizes characteristic spectral peaks to improve selectivity and enables detection of trace gases with minimal sample quantities due to the enhanced field concentration at resonant frequencies \citep{Tittl_Imaging_2018}. SEIRA spectroscopy, well-established in mid-IR chemical sensing \citep{John_Metasurface_2023}, is particularly suited for gas sensing in the mid-IR and THz regions where gases exhibit narrow absorption peaks, allowing precise identification in mixtures \citep{Gordon_The_2022}. Given that target concentrations are often in the parts-per-million (ppm) range \citep{Maleki_Detection_2015}, designing structures that reduce detection thresholds is of significant interest.
Numerous metasurfaces have been developed for SEIRA spectroscopy, utilizing either metallic nanoantennas or dielectric resonators. For metallic metasurfaces, common designs include nanorods or nanoslit structures supporting electric dipole (ED) and magnetic dipole (MD) resonances \citep{Armelles_Magnetic_2020}, and nanogap configurations that concentrate the electric field \citep{Dong_Nanogapped_2017, Brown_Fan_2015}. In all-dielectric metasurfaces, displacement currents enable the excitation of multiple resonant modes, including ED, MD, and higher-order multipoles such as electric quadrupoles (EQ) and magnetic quadrupoles (MQ) \citep{Koshelev_Dielectric_2021}. The resonance properties can be further enhanced using quasi-BICs, which, through carefully engineered asymmetry, produce high-quality, low-leakage resonances that increase both interaction time and field concentration within the resonator \citep{Bulgakov_Light_2017}. Despite the diverse resonant properties of these structures, most SEIRA metasurfaces are designed around a single resonance mode (predominantly ED) and operate in reflection mode \citep{Adato_Engineering_2015}, or transmission mode for low amplitude resonances \citep{Neubrech_Surface_2017}. However, by coupling ED and MD resonances to satisfy the Kerker condition, which minimizes backscattering, metasurfaces can achieve properties similar to Huygens sources \citep{Koshelev_Dielectric_2021, Chen_Huygens_2018}. Huygens metasurfaces, known for phase control and high transmission efficiency, have been explored for applications such as wavefront shaping \citep{Chong_Efficient_2016} and, more recently, for chiral sensing \citep{Mohammadi_Accessible_2019}. However, the use of dual resonances under the Kerker condition for SEIRA spectroscopy remains unexplored.

In this work, we present a novel free-standing metasurface composed of silicon nanodisks, designed to support quasi-BIC resonances with both ED and MD responses in the THz range. We investigate its sensitivity as a SEIRA sensor, comparing configurations based on a single resonance versus coupled ED and MD resonances that fulfill the Kerker condition. By evaluating the sensor’s performance for hydrogen cyanide (HCN) and sulfur dioxide (SO$_2$), we demonstrate how this dual-resonance approach can enhance sensitivity, making it a promising candidate for high-performance gas detection in compact sensing platforms.

\section{Metasurface design} 

\begin{figure}[h]
    \centering
    \includegraphics[width=1\linewidth]{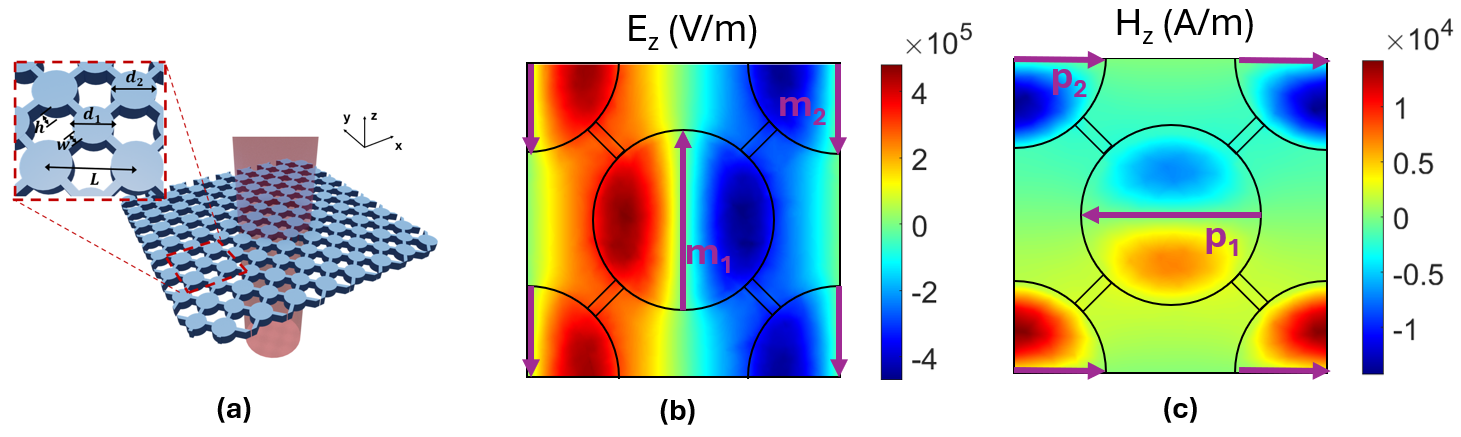}
    \caption{\textbf{ Proposed structure and electromagnetic response of the metasurface.} (a) 3D diagram of the metasurface under study. The inset shows the parameters of a unit cell. (b) Z component of the electric field for the first resonance (MD) and direction of the magnetic dipole moment $\Vec{m}$ on each resonator. (c) Z component of the magnetic field for the second resonance (ED) and direction of its electric dipole moment $\Vec{p}$ on each resonator.}
  \label{Figure1}
\end{figure}

The metasurface under study is shown in Figure \ref{Figure1}(a). It consists of an array of silicon cylinders, with a fixed distance between adjacent cylinders. There are two different possible diameters for the cylinders: $d_1$ and $d_2 = d_1 + \Delta$, where the cylinders adjacent to a cylinder with diameter $d_1$ have a diameter of $d_2$ and vice versa. The distance between two cylinders of the same diameter is $L$. All cylinders have the same height $h$. The silicon cylinders are supported by thin walls also made of silicon with a width $w$, allowing the metasurface to be free-standing, like in the case of \citep{Fan_Dynamic_2019}. This configuration is convenient for sensing gases for two main reason: First, the absence of substrate reduces the unwanted losses in the systems that can limit the performance of the metasurface as a sensor \citep{Alvarez_Loss_2023}. Second, free-standing structures, like the one proposed in this work, allow to maximize the interaction of the gas at both sides of the metasurface, improving the sensitivity. We are going to study two different modes supported by this structure, one being a magnetic dipole (MD) resonance, and the other being an electric dipole (ED) resonance. Figures \ref{Figure1}(b) and \ref{Figure1}(c) represent the field distribution of these modes when $L=226.65\, \mu m$, $h=60.5\,  \mu m$, $d_1=118.4\,  \mu m$ $d_2= 132 \, \mu m$ and $\Delta =13.6 \,  \mu m$. For all simulations in this work, we have used $n=3.42$ and $k=0.00002$ as the refractive index of silicon \citep{Dai_Terahertz_2004}.

\begin{figure}
    \centering
    \includegraphics[width=1\linewidth]{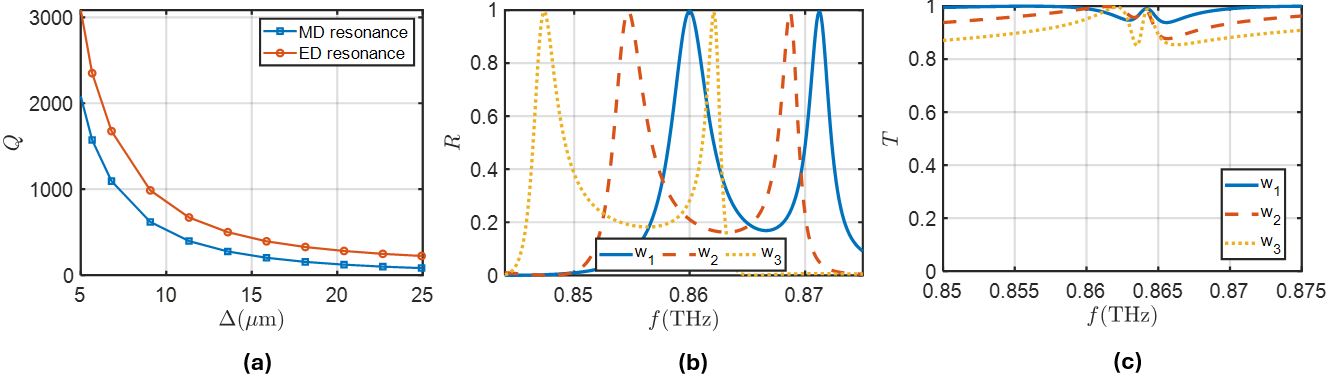}
    \caption{(a) Quality factor of both the MD and ED resonances as a function of $\Delta$. (b) Reflection spectra of the metasurface with uncoupled resonances ($L=226.65 \mu \rm m$, $d_1=118.4 \mu \rm m$, $h=60.5 \mu m$ and $\Delta =13.6 \mu \rm m$) for different values of $w$ ($w_1=2.72 \mu{\rm m}$, $w_2=10.88 \mu{\rm m}$, and $w_3=21.76 \mu{\rm m}$). (c) Transmission spectra of the metasurface with coupled resonances for different values of $w$ ($w_1=2.72 \mu{\rm m}$, $w_2=10.88 \mu{\rm m}$, and $w_3=21.76 \mu{\rm m}$). The parameters that lead to coupled resonances for each value of $w$ are: $L=231.5 \mu \rm m$, $d_1=119.7 \mu \rm m$ and $h=59.2 \mu \rm m$ for $w_1$, $L=229.6 \mu m$, $d_1=119.9 \mu \rm m$ and $h=58.8 \mu \rm m$ for $w_2$, and $L=227.7 \mu \rm m$, $d_1=119.1 \mu \rm m$ and $h=58.3 \mu \rm m$ for $w_3$, with $\Delta =13.6 \mu \rm m$ for all of them.}
    \label{fig2trip}
\end{figure}

An interesting characteristic of these modes is the way their quality factor (Q) depends on $\Delta$. When the parameter $\Delta=0$, i.e. all the nanodisks have the same diameter, the metasurface supports two different bound state in the continuum (BIC) modes that are inaccessible for normal plane-wave illumination, one of them being a magnetic dipole mode and the other an electric dipole mode. 
The origin of the BIC modes can be thought of as a superposition of two sub-lattices holding a magnetic/electric dipole moment, but with a $\pi$ relative phase between wave oscillations in the two lattices that cancels out the far-field radiation. 
In the perturbed case, when $\Delta\neq 0$, the size of the nearest neighbor changes, and the electric/magnetic dipoles induced in the nanodisks are not identical. The geometrical differences creates a difference in amplitude and/or phase that allows non-perfect destructive superposition of scattered waves in the far-field creating a high-Q resonance, know as quasi-BIC mode.
With the proposed configuration, one can not only control the position of the electric and magnetic dipole resonances but also the quality factor of the resonances. 
Figure \ref{fig2trip}(a) shows the tunability of the quality factor of these resonances with the asymmetry parameter $\Delta$, where the Q values have been obtained through an eigenmode study. This is an important feature when designing metasurfaces for enhancing absorption, allowing to tailor the scattering losses of the structure to balance the losses induced by the materials and maximize absorption.

Figure \ref{fig2trip}(b) shows the reflection spectrum of this structure for different values of the width of the supporting walls, $w$.  It can be seen that the resonances induced by the two modes appear as reflection maxima in the reflection spectrum. The presence of walls does not compromise the existence of quasi-BIC resonances because the structure maintains the same symmetry and the response remains the same as in previous works using this structure \citep{Lawrence_Nanoscale_2019,Hu_High_2020, Mañez_Extreme_2023}. However, the walls affect the quality factor of the resonances and the resonance frequency. 
In Figure \ref{fig2trip}(b), we can see that wider walls lead to higher quality factors for both resonances, although the rate at which they increase is different for each one. This means that wider walls lead to a bigger difference between the quality factors of electric and magnetic resonances.

In this work, we propose the use of a metasurface with balanced electric and magnetic dipole moments, known as the first Kerker condition. 
In absence of losses, by properly designing the parameters of the metasurface, it is possible to obtain a coupling between both resonances, which leads to the reflection peaks disappearing and having almost full transmission in the whole spectrum as seen in Figure \ref{fig2trip}(c). 
The small transmission variations near the resonance frequency are due to the fact that both resonances have different quality factor and limit the balance of electric and magnetic dipoles near the resonance. At this point it is important to notice the effect of the walls in the difference between the electric and magnetic quality factors. To minimize the perturbations in the transmission spectra we need to minimize the width of the walls and, at the same time, ensure the mechanical stability of the sample. In what follow, we will consider the width of the walls to be $w=11 \mu \rm m$ being in agreement with the fabricated samples in \citep{Fan_Dynamic_2019}. Despite the disappearance of the resonance in the frequency spectrum, there is a high field concentration at the resonance frequency that can lead to higher values of absorption if the medium surrounding the metasurface has losses. 

For a better understanding of the resonant behavior of the structure and the use for SEIRA spectroscopy, we use a coupled-mode-theory (CMT) model that will allow to understand the role of  losses in the detection. This scattering of this metasurface can be modeled as a system with two ports and two resonances that, following similar approach than \citep{Suh_Temporal_2004}, give as the following reflection and transmission coefficient:
\begin{eqnarray}
    T=t_{\rm d}- \frac{\gamma_{\rm m} (r_{\rm d}+t_{\rm d})}{j(\omega -\omega_{\rm m})+\gamma_{\rm m} + 1/\tau_{\rm m}}+\frac{\gamma_{\rm e} (r_{\rm d}-t_{\rm d})}{j(\omega -\omega_{\rm e})+\gamma_{\rm e} + 1/\tau_{\rm e}} \\
    R=r_{\rm d}- \frac{\gamma_{\rm m} (r_{\rm d}+t_{\rm d})}{j(\omega -\omega_{\rm m})+\gamma_{\rm m} + 1/\tau_{\rm m}}-\frac{\gamma_{\rm e} (r_{\rm d}-t_{\rm d})}{j(\omega -\omega_{\rm e})+\gamma_{\rm e} + 1/\tau_{\rm e}} 
\label{absorption}
\end{eqnarray}
where $r_{\rm d}$ and $t_{\rm d}$ are the direct reflection and transmission coefficients, $\omega _{\rm m,e}$ and $\gamma _{\rm m,e}$ are the resonance frequency and decay rate for the magnetic and electric resonances, and the material losses are modeled by $1/\tau _{\rm m,e}$  \citep{Mañez_Extreme_2023}
Finally, an equation for the absorption spectrum can be reached with coupled mode theory, $A=1-|R|^2-|T|^2$. Using that the reflection outside of the resonances is almost null, we can consider $r_d = 0$ which simplifies the equation to:

\begin{equation}
    A=1-t_{\rm d}^2 \left(1- \frac{2\gamma_{\rm e}\tau_{\rm e}}{1+\tau_{\rm e}(2\gamma_{\rm e}+\gamma_{\rm e}^2\tau_{\rm e}+\tau_{\rm e}(\omega-\omega_{\rm e})^2)}-\frac{2\gamma_{\rm m}\tau_{\rm m}}{1+\tau_{\rm m}(2\gamma_{\rm m}+\gamma_{\rm m}^2\tau_{\rm m}+\tau_{\rm m}(\omega-\omega_{\rm m})^2)}\right)
\label{absorption_simplified}
\end{equation}

\begin{figure}[h]
    \centering
    \includegraphics[width=1\linewidth]{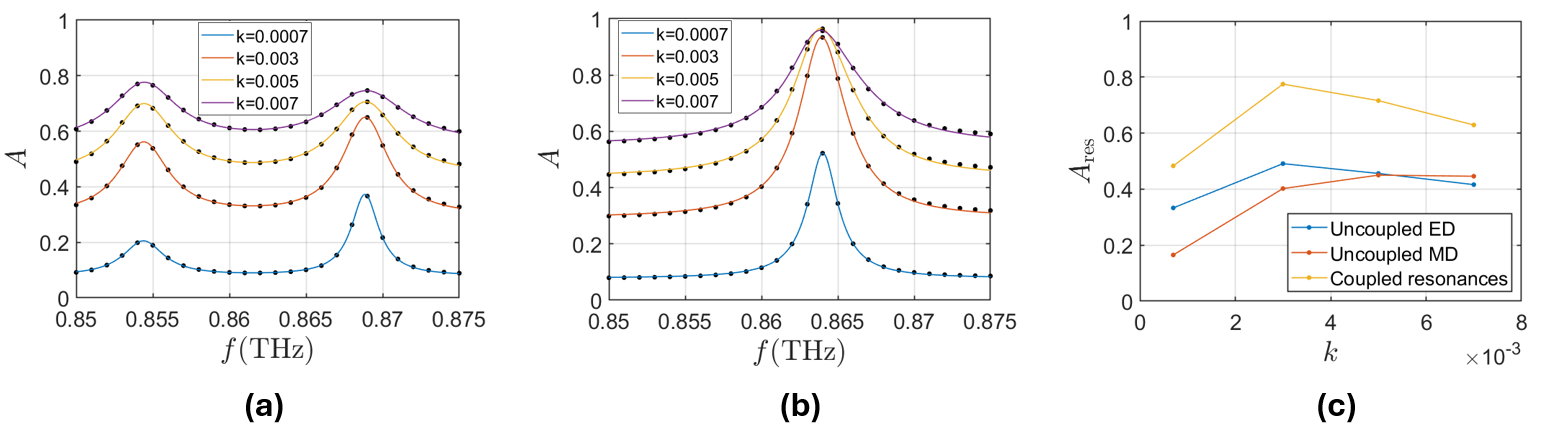}
    \caption{Absorption spectra of the uncoupled (a) and coupled (b) resonances for different values of the extinction coefficient of the medium. The parameters of the metasurface are $L=226.65 \mu \rm m$, $d_1=118.4 \mu \rm m$, $h=60.5 \mu \rm m$ for the uncoupled case and $L=229.6 \mu \rm m$, $d_1=119.9 \mu \rm m$, $h=58.8 \mu m$ for the coupled case, with $w=10.88 \mu \rm m$ and $\Delta =13.6 \mu \rm m$ for both. (c) $A_{\rm res}$ as a function of $k$ for both the uncoupled and the coupled resonances.}
    \label{figure3}
\end{figure}

Figure \ref{figure3}(a) shows the absorption spectrum obtained through simulation and its fit to Eq. \ref{absorption} for the case with uncoupled, $\omega_{\rm e}\neq\omega_{\rm m}$ for different values of the losses of the background media where the metasurface is suspended, while Figure \ref{figure3}(b) shows the same for the coupled resonances working under Kerker condition ($\omega_{\rm e}=\omega_{\rm m}$), where the structure has been designed to keep the similar quality factors for both resonant mode to facilitate the comparison with the previous example. Table \ref{tab:fit} shows the parameters of the fits to Eq. \ref{absorption}. From this analysis, we can see that, as expected, the parameters related to the resonances ($\omega _{m,e}$ and $\gamma _{m,e}$) do not depend on $k$ of the medium. As for the material losses, $1/\tau_{m,e}$, they increase with higher values of $k$, while the transmission of the direct channel, $t_{\rm d}$, that emulates the dissipation of energy due to free wave propagation, decreases with higher values of $k$. We can define absorption due to the resonance (without the effect of the lowered direct transmission) as $A_{\rm res}= A(\omega_{\rm m,e})-(1-t_{\rm d})$, the maximum of which is obtained for each resonance when $1/\tau_{m,e}=\gamma _{m,e}$. If further material losses are added, this value of absorption decreases, as seen in Figure \ref{figure3}(c). For the uncoupled ED resonance and for the two resonances case, this maximum is obtained for $k=0.003$, while for the uncoupled ED resonance it would be obtained at higher values of $k$.
This limit will define the maximum amount of losses that can be detected and it is closely related with the maximum concentration of gas that can be introduced into the system as we will see in the next section. The results show that, in the uncoupled case, the maximum absorption does not reach unity with only one resonance dissipating energy due to the fact that total absorption requires the simultaneous interaction of electric and magnetic resonances \citep{Radi_Total_2013}, or, in other words, working under the Kerker condition. In the coupled case, where the Kerker condition is met, the maximum absorption reaches values close to 1. 

With this configuration the metasurface will work on transmission mode, in contrast with the previous example where the resonances produced high reflection. The amplitude of the transmission coefficient will be defined between zero and unity improving the sensitivity of the metasurface with a single resonance where the reflection coefficient has half span. Another aspect to discuss in this structure is the effect of the optical path surrounding the metasurface that defines the out-of-resonance absorption  (defined by the parameter $t_{\rm d}$ in our model). The larger the optical path surrounding the metasurface is, the smaller absorption span.  For this reason, it is preferable to work with short optical paths.

\begin{table}[h]
    \centering
    \caption{Parameters of equation (\ref{absorption}) for the fitted absorption curves. The parameters of the metasurface are $L=226.65 \mu \rm m$, $d_1=118.4 \mu \rm m$, $h=60.5 \mu \rm m$ for the uncoupled case and $L=229.6 \mu \rm m$, $d_1=119.9 \mu \rm m$, $h=58.8 \mu m$ for the coupled case, with $w=10.88 \mu \rm m$ and $\Delta =13.6 \mu \rm m$ for both.}
    \begin{tabular}{|c|c|c|c|c|c|c|c|}
    \hline
         & $\omega _m $ [THz] & $\omega _e$ [THz] & $\gamma _m$ [THz] & $\gamma _e$ [THz]& $1/\tau _m$ $[\rm THz]$ & $1/\tau _e$ $[\rm THz]$& $t_d$\\
         \hline
         Uncoupled, $k = 0.0007$& 0.8544 & 0.8688 & 0.0015 & 0.0009 & 0.0001 & 0.0002 & 0.9604 \\
         \hline
         Uncoupled, $k = 0.003$& 0.8544 & 0.8689 & 0.0015 & 0.0009 & 0.0005 & 0.0009 & 0.8413 \\
         \hline
          Uncoupled, $k = 0.005$& 0.8544 & 0.8689 & 0.0015 & 0.0009 & 0.0008 & 0.0015 & 0.7499 \\
         \hline
          Uncoupled, $k = 0.007$& 0.8544 & 0.869 & 0.0016 & 0.0009 & 0.0011 & 0.0022 & 0.6687 \\
         \hline
         Coupled, $k = 0.0007$& 0.864 & 0.864 & 0.0014 & 0.0009 & 0.0002 & 0.0002 & 0.9604 \\
         \hline
         Coupled, $k = 0.003$& 0.8643 & 0.8638 & 0.0015 & 0.0008 & 0.0007 & 0.0008 & 0.8418 \\
         \hline
         Coupled, $k = 0.005$& 0.8644 & 0.8637 & 0.0017 & 0.0007 & 0.0012 & 0.0011 & 0.7512 \\
         \hline
         Coupled, $k = 0.007$& 0.8646 & 0.8637 & 0.0012 & 0.0011 & 0.0026 & 0.0011 & 0.6719 \\
         \hline
         
    \end{tabular}

    \label{tab:fit}
\end{table}

\newpage

\section{Metasurface use for gas sensing} 

\begin{figure}[b]
\centering
\includegraphics[width=1\linewidth]{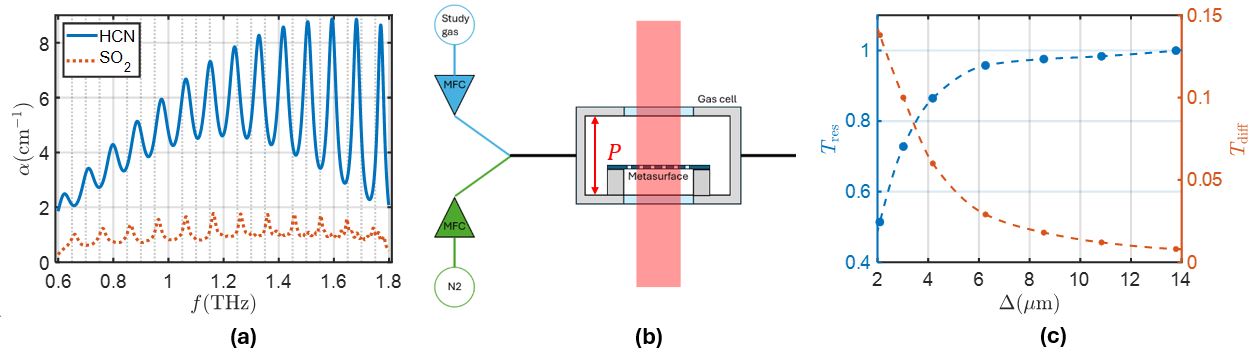}
\caption{(a) Absorption coefficient spectrum of $\rm HCN$ and $\rm SO_2$. (b) Scheme of gas sensing setup. (c) Transmission at the resonance without gas (blue curve) and transmission difference at the resonance for 1000 ppm of $\rm SO_2$ (orange curve) as a function of $\Delta$ (see Appendix for the metasurface parameters for each value of $\Delta$).
\label{fig4}}
\end{figure}

To test the performance of this structure as a sensor, we will center our study on two sample gases: $\rm HCN$ and $\rm SO_2$. $\rm HCN$ is an extremely poisonous gas that is used in several industrial processes. It has distinctive fingerprints in the THz band, as seen in Figure \ref{fig4}(a). In particular, we will use the absorption peak at 1.5 THz to sense the gas concentration.
$\rm SO_2$, while not as dangerous as $\rm HCN$, is also used in industrial processes and it can be interesting to measure its concentration with precision. It also has absorption peaks in the THz region, although they are not as intense as the ones of $\rm HCN$, as seen in Figure \ref{fig4}(a). We will focus on the absorption peak at 0.86 THz, since it is the strongest absorption peak that does not overlap with water vapor peaks.
In the simulations, the gas under study will be a mix of one of these gases with $\rm N_2$, since it is the predominant gas in air and it has no absorption peaks in the THz range. 
The numerical study performed in this section will evaluate the performance of the metasurface as a SEIRA spectroscopy sensor with the setup configuration shown in Fig. \ref{fig4}(b).
The metasurface is placed inside a gas cell with an optical path $P$, which in the case of our simulations is 3 mm. The gas concentration could be adjusted  using two mass flow controllers (MFC) to achieve the desired gas concentrations.

For a given concentration of gas in the environment, the geometrical parameters of the metasurface can be chosen so that the coupled resonance appears at the same frequency and the losses induced in the materials and the scattering losses are balanced to achieve the condition of critical coupling that ensures full absorption of incoming energy at the resonance frequency. 
This way, the transmission amplitude at the resonance frequency can be used to determine the concentration of the target gas in a mixture with other gases that do not have absorption at the same frequency.
It is important to note that, thanks to the capability of controlling the quality factor of the resonances, we can control the sensitivity of the sensor and adjust the full absorption condition to the specific concentration that is targeted by the application. 
As it was shown in Fig. \ref{fig2trip}(a), the proposed metasurface allows to control the quality factor of both electric and magnetic by controlling the asymmetry parameter, $\Delta$, and can achieve, at least theoretically extremely high quality factor that could lead to ultra-sensitive sensors.
To see how the sensitivity changes with the value of $\Delta$, we simulated the transmission spectra for 1000 ppm of $\rm SO_2$ and obtained a transmission difference ($T_{\rm diff}$) between the transmission at the resonance with and without gas [see Figure \ref{fig4}(c)].
Tuning the values of $\Delta$, we can increase the sensitivity while still having a resonance that can be observed with current THz systems.


However, there are two main limiting factors to the quality factor that is attainable by our metasurface. The first is that, despite being very low, the material losses of the silicon resonators have an effect on the transmission amplitude of the resonance \citep{Alvarez_Loss_2023}. This effect as a function $\Delta$ can be seen in Figure \ref{fig4}(c). The second factor is the spectral resolution of the THz systems. If a metasurface with a high enough quality factor was fabricated, the resonance would not be visible using current THz systems. The FWHM of the resonance has to be several times bigger than the resolution of the system so that the transmission peak at the resonance is well defined and can be properly used for sensing. At THz frequencies, using time-domain spectroscopy, the resolution is in the range between 0.5 GHz and 1 GHz \citep{Tspec,K15,TeraFlash}. This resolution can be improved using frequency domain spectroscopy up to 0.1 GHz \citep{TerasScan, Kong_High_2018, Vogt_Coherent_2019,verTera}.  In what follows, these two aspects will be considered in choosing the asymmetry parameter.
For example, we estimate that, for sensing the 0.864 THz $\rm SO_2$ resonance, a quality factor lower than 8600 could be measured experimentally and, with the structure dimensions studied in the previous section, a value of $\Delta>2\mu$m guarantees this criterion [see Fig.\ref{fig2trip}(a)], while for the 1.505 THz $\rm HCN$ resonance a quality factor lower than 15000 would be measurable, which corresponds to $\Delta>1.2\mu$m. With this criterion in mind, two different structures have been designed to detect $\rm HCN$ and $\rm SO_2$.

\begin{figure}[h]
\centering
\includegraphics[width=1\columnwidth]{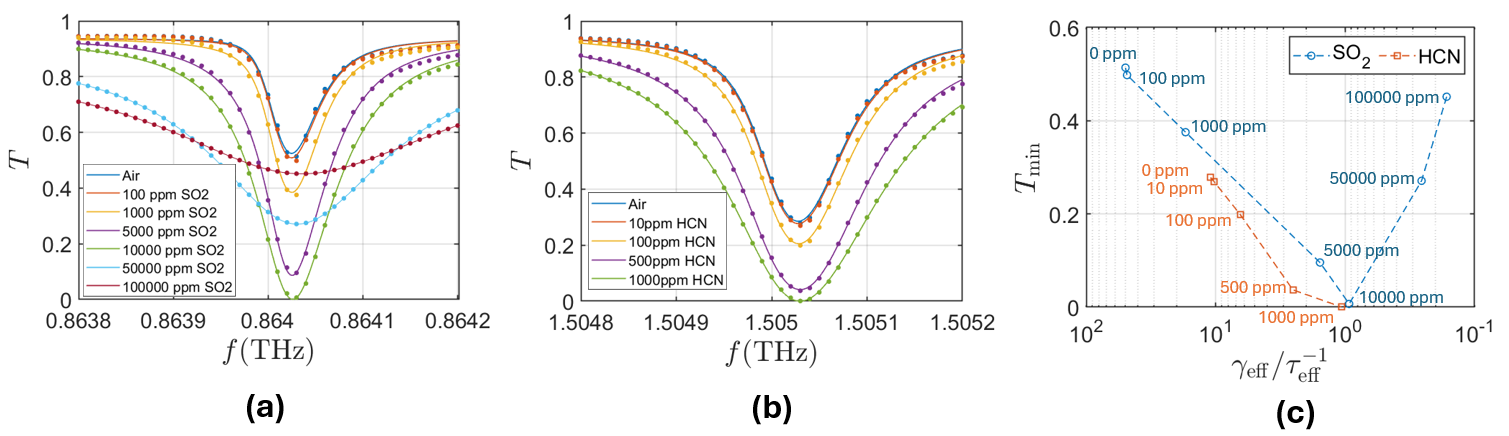}
\caption{Transmission spectra of the high Q coupled resonance for different values of (a) $\rm SO_2$ concentration, and (b) $\rm HCN$ concentration, with $L=232.35 \mu m$, $d_1=127.26 \mu m$, $h=58.21 \mu m$ and $\Delta =2.09 \mu m$ for the $\rm SO_2$ case and $L=133.39 \mu m$, $d_1=73.06 \mu m$, $h=33.42 \mu m$ and $\Delta =1.2 \mu m$ for the $\rm HCN$ case. (c) Transmission minimum as a function of the ratio between $\gamma_{\rm eff}$ (average of $\gamma_{\rm e}$ and $\gamma_{\rm m}$) and $1/\tau_{\rm eff}$ (average of $1/\tau_{\rm e}$ and $1/\tau_{\rm m}$) for both metasurfaces, i.e. the ratio between the radiative losses of the resonance and the losses due to absorption from the surrounding medium.
\label{fig5}}
\end{figure}

Figure \ref{fig5}(a) shows the coupled resonance transmittance for $L=232.35 \mu m$, $d_1=127.26 \mu m$, $h=58.21 \mu m$ and $\Delta =2.09 \mu m$ for different concentrations of $\rm SO_2$. As we previously saw in Figure \ref{fig4} (c), the resonance no longer reaches values near full transmission due to the losses of silicon. However, the difference in transmission when adding the gas is larger than for lower values of Q. We can also observe that for concentrations higher than 10000 ppm, the transmission at the resonance increases due to the effect of losses. This means that there is an optimum value of losses in the system that, when exceeded, leads to less absorption.\citep{Alvarez_Loss_2023}. As seen in Figure \ref{fig5}(c) , this maximum value of absorption is achieved for $1/\tau_{\rm eff} = \gamma_{\rm eff} $, where we have defined $\tau_{\rm eff}$ and $\gamma_{\rm eff}$ as the average of $\tau_{\rm m, e}$ and $\gamma_{\rm m,e}$, respectively. However, even if the same value of transmission minimum is reached for two different values of concentration, they could still be distinguished because of their different transmission values outside the resonance frequency. 
To assess the performance of our metasurface, we compare it to the free space transmission values. Using the Beer-Lambert law, $I=I_0 e^{-\alpha \cdot L}$, we can obtain the transmission as a function of the optical path length through different concentrations of $\rm SO_2$ and $\rm HCN$. From this, we can compare the path length needed to obtain the same difference in transmission between the system with a metasurface and the free space measurement, as shown in Table \ref{tab:Table2}. Overall, the path length ratio between the use of this metasurface, which requires a path length of 0.15 mm, and free-space is between 400 and 3700, depending on the concentration of the gas. The smaller the concentration of gas being measured, the more efficient our system is compared to free space sensing.

 \begin{table}[!ht]
 \centering
  \caption{Transmission difference between system with air and with each value of concentration and equivalent path length in free space measuring, as well as the ratio between this path length and the minimum path length for the metasurface (0.15 mm)}
 \begin{tabular}{|l|l|l|l|}
 \hline
    Gas concentration                    & $T_{\rm diff}$    & $P_{\rm free \, space}$ &   $P_{\rm ratio}$\\
                        \hline
 100 ppm SO$_2$        & 0.016    & 56.5 cm &  3767 \\
 \hline
 1000 ppm SO$_2$       & 0.139      & 52.5 cm & 3500 \\
 \hline
 5000 ppm SO$_2$    & 0.418     & 38.3 cm &  2553 \\
 \hline
 10000 ppm SO$_2$       & 0.514 & 25.8 cm&  1720 \\
 \hline
 10 ppm HCN         & 0.009 & 17.1 cm &  1140 \\
 \hline
 100 ppm HCN         & 0.079 & 15.5 cm &  1033 \\
 \hline
 500 ppm HCN           & 0.241 & 10.4 cm &  693 \\
 \hline
 1000 ppm HCN          & 0.278 & 6.2 cm & 413 \\
 \hline
 \end{tabular}

 \label{tab:Table2}
 \end{table}

To compare our metasurface with other gas sensing systems, we study an scaled version of our metasurface so that the resonance appears at 1.5 THz and see the effect of $\rm HCN$ on the resonance. Figure \ref{fig5}(b) shows the transmission spectrum as a function of $\rm HCN$ concentration for this metasurface, the parameters of which are $L=133.39 \mu m$, $d_1=73.06 \mu m$, $h=33.42 \mu m$ and $\Delta =1.2 \mu m$. In Table \ref{tab:Table3}, we compare our system to other THz gas sensing devices, both in terms of path length needed and minimum concentration of $\rm HCN$ detectable. For our system, this minimum would be 10 ppm of HCN if we consider a difference  in the transmission spectrum $T_{\rm diff}^{\rm min}\approx 0.01$ to be the detection threshold. After several simulations reducing the path length, we reached the conclusion that the minimum path length needed to avoid losing sensitivity is $P_{\rm min}=0.15$ mm. According to the simulations, our system is the one that would require the shorter path length, although it would not reach as low a threshold as other systems. We can define a figure of merit $FOM={1}/{(P_{\rm min} \cdot T_{\rm diff}^{\rm min})}$ that takes into account both the detection threshold and the path length of the detection system. Using this figure of merit, we can see that the system presented in this work obtains similar values to the most sensitive THz gas sensors.

Another benefit from our system, apart from the high sensitivity, is that it works in transmission. This, coupled with the fact that the path length requisite is small, would allow us to stack several metasurfaces to improve the functionalities of our sensor while only increasing the path length. If the metasurfaces stacked are identical, we would get a higher sensitivity for the target gas. However, stacking metasurfaces with different parameters could also have interesting applications, due to the transmission outside the resonance being close to 1. If the resonances are equally spaced in a certain frequency band, the molecular fingerprint of the gas mixture in this band could be obtained \citep{Tittl_Imaging_2018}. If the metasurfaces are designed to have their resonances at the same frequency as the absorption peak of different gases, we would obtain a multi gas sensor.

 \begin{table}[!ht]
 \centering
  \caption{Comparison of the path length needed, the detection threshold of HCN concentration and the figure of merit between different THz gas sensors}
 \begin{tabular}{|l|l|l|l|}
 \hline
                        & Path length    & Threshold &   FOM\\
                        \hline
 Photonic crystal cavity \citep{Shi_Highly_2018}        & 5 mm      & 2 ppm &  0.1 $\rm mm^{-1} ppm^{-1}$ \\
 \hline
 Photonic crystal fiber \citep{Qin_Terahertz_2019}        & 100 mm      & 2 ppm & 0.005 $\rm mm^{-1} ppm^{-1}$ \\
 \hline
 Resonant cavity \citep{Elmahle_THz_2023}  & 480 mm     & 0.003 ppm &  0.694 $\rm mm^{-1} ppm^{-1}$ \\
 \hline
 This work         & 0.15 mm & 10 ppm &  0.667 $\rm mm^{-1} ppm^{-1}$ \\
 \hline
 \end{tabular}

 \label{tab:Table3}
 \end{table}


\section{Conclusions} 
In this work, we have studied a terahertz gas sensor based on an all-dielectric metasurface composed of Si nanodisks with thin walls between them. This structure supports two QBIC resonances that can be coupled to satisfy the Kerker condition, leading to increased interaction with the surrounding gas and the possibility of performing measurements in transmission mode. This effect, coupled with the lack of substrate, the Q factor tunability of the resonances and the low losses of Si in the THz range, makes this a promising structure for the development of gas sensors. Through the use of coupled mode theory, an expression for the absorption spectrum of the system has been obtained and used to fit the simulation data. From this model we can predict the the existence of an optimal amount of losses from the medium surrounding the metasurface, which will mark the maximum gas concentration detectable by the system.

We then study the sensitivity of this system for binary mixtures of $\rm N_2$ and one of two different gases, $\rm SO_2$ and $\rm HCN$, both of which have some intense absorption peaks in the THz range. If we compare the minimum path length needed for the metasurface to reach maximum sensitivity with free space sensing, we see that the metasurface requires a path between 3700 and 400 times shorter, with our system being more efficient the smaller the gas concentration under measurement is. This opens a path to develop ultra-compact gas sensors for industrial system where the availability of space is severely constrained as inside industrial or chemical production machinery. When comparing the $\rm HCN$ case to other THz gas sensing systems not based on free space sensing, the results obtained are that the detection threshold is higher than that of other systems, but the path length needed is much shorter. For better comparison with other systems, we have defined a new figure of merit that takes into account both the minimum amount of gas measurable as well as the length of the optical path needed to detect this gas concentration. The values of this figure of merit obtained by our sensor are comparable to some of the most sensitive gas sensing systems.  

 Future research could explore the possibility of stacking several metasurfaces for enhanced sensitivity or multi-gas detection, taking advantage of the fact that the system works in transmission, as well as combining it with other gas sensing THz systems. These advancements could solidify the role of this technology in high-precision, real-time gas sensing applications, specially in those where the space that can be used for sensing is limited.

\section*{Acknowledgements}

This work was supported in part by the \href{http://dx.doi.org/10.13039/501100004837}{\underline{Ministerio de Ciencia
e Innovación}} –\href{http://dx.doi.org/10.13039/501100011033}{\underline{Agencia Estatal de Investigación}} under Project PID2019-
111339GBI00 and TED2021-132259B-I00. Action co-financed by the European
Union through the \\
\href{http://dx.doi.org/10.13039/501100008530}{\underline{ European Regional Development Fund (ERDF)}}
operational program for the Valencian Community 2014-2020.

\section*{Author contributions statement}

All authors have accepted responsibility for the entire content of this manuscript and approved its submission.

\section*{Competing interests}
The authors declare no competing interests.

\newpage

\section*{Appendix: Parameters for Figure \ref{fig4}(c)}

\begin{table}[h]
\centering
\begin{tabular}{|c|c|c|c|}
\hline
  $\Delta$ & $L$    & $d_1$ &  $h$\\
                       \hline
2.09 $\mu$m       & 232.35 $\mu$m      & 127.26 $\mu$m & 58.21 $\mu$m \\
\hline
3.02 $\mu$m       & 232.23 $\mu$m      & 126.73 $\mu$m & 58.21 $\mu$m \\
\hline
4.18 $\mu$m       & 232.13 $\mu$m      & 126.10 $\mu$m & 58.24 $\mu$m \\
\hline
6.26 $\mu$m       & 231.86 $\mu$m      & 124.90 $\mu$m & 58.33 $\mu$m \\
\hline
8.56 $\mu$m       & 231.27 $\mu$m      & 123.43 $\mu$m & 58.41 $\mu$m \\
\hline
10.83 $\mu$m       & 230.58 $\mu$m      & 121.91 $\mu$m & 58.54 $\mu$m \\
\hline
13.60 $\mu$m       & 229.60 $\mu$m      & 119.90 $\mu$m & 58.80 $\mu$m \\
\hline
\end{tabular}
\label{tab:Tableapp}
\end{table}
\end{document}